\documentclass[reprint, 10pt, a4paper, aps, prx, floatfix, longbibliography]{revtex4-1}
\usepackage[utf8]{inputenc}
\usepackage{lipsum}
\usepackage{graphicx}
\usepackage{dcolumn}
\usepackage{bm}
\usepackage{amssymb}
\usepackage{amsmath}
\usepackage{textcomp}
\usepackage{color}
\usepackage{mathtools}

\begin{document}

\title{Suppression of power losses during laser pulse propagation in underdense plasma slab}

\author{K. V. Lezhnin}
\email{klezhnin@princeton.edu}
\author{K. Qu}
\author{N. J. Fisch}
\affiliation{Department of Astrophysical Sciences, Princeton University, Princeton, New Jersey 08544, USA}

\date{\today}

\begin{abstract}
For current state-of-the-art terawatt lasers, the primary laser scattering mechanisms in plasma include Forward Raman Scattering (FRS), excitation of plasma waves, and the self-modulational instability (SMI). Using 2D PIC simulations, we demonstrate the dominance of the FRS in the regime with medium-to-low density plasma and non-relativistic laser fields. However, the use of multi-colored lasers with frequency detuning exceeding the plasma frequency, $\Delta \omega >\omega_{\rm pe}$, suppresses the FRS. The laser power can then be transmitted efficiently.

\bigskip

\end{abstract}

\maketitle

\section{Introduction}

In recent years, high power laser technology has reached the level of petawatt scale with kilojoule laser pulse energy \cite{PETAWATTREVIEW}. High power laser-matter
interactions result in the generation of high energy beams
of charged particles, electrons \cite{LWFA1,LWFA2} and ions \cite{LIA}, photons in a
wide frequency range spanning from low-frequency electromagnetic pulses to X-rays \cite{XRAY} and $\gamma$-rays \cite{GAMMARAY}, offering a broad range of possible particle/light sources with numerous applications \cite{APPLICATIONS}. 

In applications of laser-matter interactions, it is vital to transmit peak laser power to the target without incurring substantial energy loss. For example, laser-ion acceleration requires the propagation of high laser power in underdense plasma to achieve high efficiency~\cite{Esirkepov2014,Lezhnin2016}. Similarly, the generation of $\gamma$-rays through laser-solid interactions \cite{Nakamura2012,Lezhnin2018} requires propagation at high laser intensity. In both applications, the unwanted laser scattering happens mainly due to Forward Raman Scattering (FRS)~\cite{FRS} and filamentation instabilities~\cite{Max1974,FilamentationReview}. Yet another application requiring laser transmission at high power is a plasma-based laser amplifier, working through either Backward Raman Scattering~\cite{MF1999} or Stimulated Brillouin Scattering~\cite{SBS}. Along with other effects, both the FRS and filamentation instabilities {can interfere with this transmission} \cite{Trines2010}.

Several methods were explored and have demonstrated success in avoiding these instabilities: introducing a second laser with a slight frequency shift of the order of plasma frequency~\cite{Kalmykov2006,Kalmykov2008,Kalmykov2009}; separating the total laser power into multiple subcritical laser pulses with the controlled coalescence of these pulses~\cite{Askaryan1994,Fairchild2017}; and introducing spatial incoherence of the laser pulse in order to avoid the critical power for the filamentation instability~\cite{VM2016}. 

Here we focus on the method of frequency detuning, where two detuned copropagating laser pulses can suppress both FRS and relativistic filamentation. The ponderomotive potential of the laser beat drives a plasma density modulation which can either enhance or suppress the instabilities depending on the laser frequency detuning, $\Delta\omega$. Specifically, our fully relativistic, kinetic PIC simulations demonstrate that, by using two pulses with detuning $\Delta \omega/\omega_{\rm pe}> 1$, laser power is propagated more efficiently owing to suppression of both the FRS and filamentation instabilities. 
We also show the optimal energy partition of the frequency components for the maximum pulse power transmission.

Our finding supplements the findings by Kalmykov, et al~\cite{Kalmykov2006, Kalmykov2008, Kalmykov2009}, namely that the two-color laser scheme with $\Delta\omega > \omega_{\rm pe}$ can avoid catastrophic relativistic filamentation within a propagation distance of a few Rayleigh lengths. Although their main interest is the suppression of relativistic self-focusing, their numerical simulations exhibit tail erosion due to FRS and electromagnetic cascades. The tail refers to the less intense, off-peak, part of a laser beam. Here we point out that, for parameters of interest, the most significant power loss is in fact caused by the slow-moving Stokes sidebands which temporally separate from the main pulse. Filamentation only modulates the laser pulse envelope. Thus, demonstrating the ability to suppress FRS broadens the applicability of the two-color laser scheme in long-distance high-power electromagnetic power propagation.

The paper is organized as follows. In Section II, we review the basic theoretical concepts and distinguish the dominant power loss mechanism. In Section III, we describe the setup of the numerical simulations. In Section IV we discuss the simulation results and interpret them using simple theoretical estimates. In Section V, we compare our findings to previous descriptions of the detuning effect and summarize our results.

\section{Theoretical background}

An intense laser pulse, propagating in homogeneous rarefied plasma, 
drives plasma electrons to near relativistic speeds, thereby increasing the electron mass and decreasing the plasma frequency. The intensity-dependent change of the refractive index focuses the laser pulse, leading to relativistic self-modulation. The growth rate of the self-modulational instability (SMI)~\cite{Max1974,FilamentationReview} is
\begin{equation}
    \gamma_{\rm SMI,max} T_0 =  \frac{\pi}{4} a_0^2 \frac{\omega_{\rm pe}^2}{\omega_0^2}\frac{1}{(1+a_0^2)^{3/2}},
    \label{smrate}
\end{equation}
where $\omega_0 = 2 \pi /T_0$ is the laser frequency in vacuum, $T_0 = \lambda/c$ is the laser period, $\lambda$ is the laser wavelength in vacuum, $c$ is the speed of light in vacuum, and $\omega_{\rm pe}=(4 \pi n_e e^2/m_e)^{1/2}$ is the plasma frequency of plasma with electron number density $n_e$. $e$ and $m_e$ denote the electric charge and mass of the electron, respectively. $a_0 =eE/m_e \omega_0 c = 0.85 \cdot \sqrt{I/10^{18} \,{\rm W/cm^2}} \cdot \lambda / 1 \rm \, \mu m$ is the dimensionless amplitude of the laser pulse. The SMI growth rate is the same for longitudinal and transverse modes. However, the fastest growing modes of the instability depend on the direction of the mode, $k_{\rm SMI,\parallel} \lambda = \pi a_0$ for the mode parallel to the laser wave vector, and $k_{\rm SMI,\perp}  \lambda = \pi a_0 {\omega_{\rm pe}}/{\omega_0}$ for the mode perpendicular to the laser wave vector. Another figure of merit to describe the growth of SMI is the e-folding number $N_{\rm e, SMI} = \gamma_{\rm SMI, max} T_{\rm int}$ with $T_{\rm int}$ being the interaction time. For a finite width, a laser pulse with power greater than the critical power $P_\mathrm{cr,rel}$ would catastrophically focus itself into filaments. The commonly accepted critical power for the relativistic filamentation obtained both numerically and analytically \cite{FilamentationReview} is
\begin{equation}
    P_{\rm cr,rel} = 17\, {\rm GW}\, \cdot \frac{\omega_{0}^2}{\omega_{\rm pe}^2}.
    \label{pcr}
\end{equation}
For the current study, we ignore other sources of the filamentation, such as ponderomotive and thermal filamentation. The relativistic filamentation dominates for terawatt pulses on a few tens of picoseconds timescale~\cite{Li2019}, though these filamentation instability branches have smaller power thresholds. 

Another major mechanism of power loss from the primary pulse is the development of the Forward Raman Scattering (FRS). It leads to the decay of the laser wave into plasma wave and a new electromagnetic wave with the wavevector
\begin{equation}
    k_{\rm FRS} = \frac{\omega_0}{c} \sqrt{\left(1-\frac{\omega_{\rm pe}}{\omega_0}\right)^2-\frac{\omega_{\rm pe}^2}{\omega_0^2}}.
    \label{kFRS}
\end{equation}
With sufficiently large amplitude and long interaction time, higher order FRS could appear at wavevectors $k_{\rm FRS}^{(p)}$ with order number $p$
\begin{equation}
    k_{\rm FRS}^{(p)} = \frac{\omega_0}{c} \sqrt{\left(1-p \cdot \frac{\omega_{\rm pe}}{\omega_0}\right)^2-\frac{\omega_{\rm pe}^2}{\omega_0^2}}.
    \label{knFRS}
\end{equation}
 Negative integer values of $p$ will correspond to the anti-Stokes components of the FRS. The growth rate of the instability is \cite{FRS}
\begin{equation}
    \gamma_{\rm FRS} T_0 =  \frac{\pi}{2} \frac{\omega_{\rm pe}^{3/2}}{\omega_0^{3/2}} \frac{a_0}{(1+a_0^2)^{7/4}}.
    \label{FRSrate}
\end{equation}
As we will show, this high-order FRS instability is observed in our simulations since the e-folding number $N_{\rm e, FRS} \equiv 2 \gamma_{\rm FRS} (T_{\rm int}\tau_L)^{1/2}$ \cite{Esarey2009} ($\tau_L$ is the laser pulse duration) exceeds $10$ at the end of interaction.

\begin{figure}
    \centering
    \includegraphics[width=\linewidth]{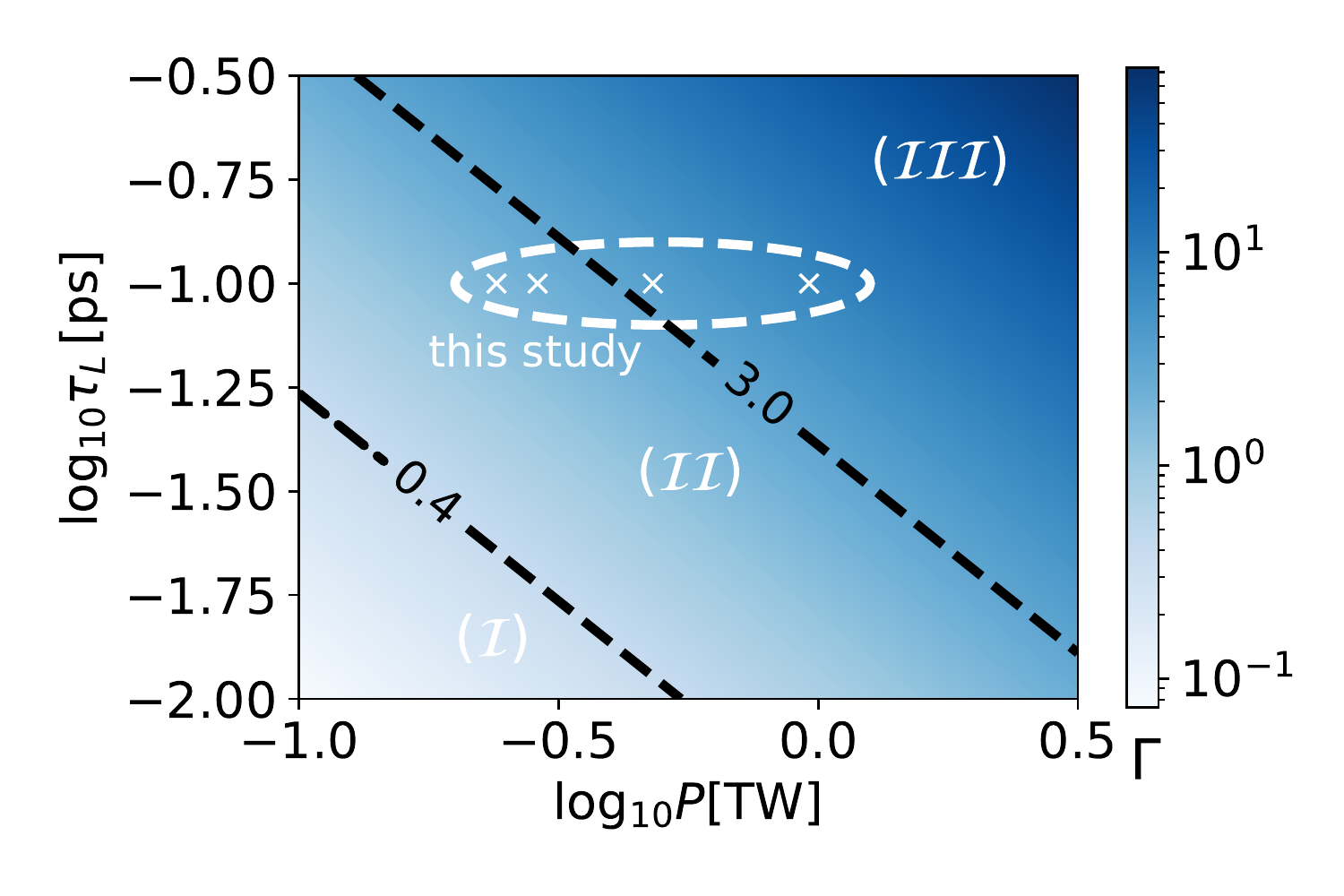}
    \caption{Dependence of $\Gamma$ on laser pulse power and duration for $n_e/n_{\rm cr}=0.05$ and $\lambda=1\, \mu \rm m$. The dashed black lines denote boundaries of regimes with $(\mathcal{I})$ SMI domination, $(\mathcal{II})$ SMI and FRS competing, and $(\mathcal{III})$ FRS domination. The white crosses demarcate our simulations parameters.}
    \label{fig:growthrates}
\end{figure}

{While the straightforward comparison of the growth rates (Eqns. (1) and (5)) correctly captures the dominant instability at the early times of the interaction, the asymptotic integration of the laser envelope evolution equation is required to reproduce the physics at later stages of envelope evolution \cite{FRS}.} We categorize the laser parameters into different regimes by invoking the following criterion (see Eq. (71)a from \cite{FRS}) based on the parameter 
\begin{equation}
    \Gamma \equiv \frac{P}{1 \, \rm TW} \cdot \frac{\tau_L}{1 \, \rm ps} \cdot  \left(\frac{n_{e}}{10^{19}\, \rm cm^{-3}} \right)^{5/2} \cdot \left( \frac{\lambda}{1 \, \rm \mu m} \right)^{4} .
\end{equation}
FRS dominates if $\Gamma \geq 3$, and SMI dominates if $\Gamma \leq 0.4$. We choose our simulation parameters to span over different interaction regimes where either SMI and FRS compete or FRS dominates, as shown through the white crosses in Fig.~\ref{fig:growthrates}. Thus, we expect the FRS to be the most dominant instability in our runs, while SMI would be a secondary factor. Our simulations supplement those reported in \cite{Kalmykov2006, Kalmykov2008, Kalmykov2009} which solely focus on the SMI-dominant regime.

\section{Simulation setup}
We perform 2D particle-in-cell (PIC) simulations using the code EPOCH \cite{EPOCH}. We considered $\tau=100$ fs Gaussian laser pulses with $w_0=20 \, \mu \rm m$ waist and linear polarization ($E_z$ is out of simulation plane $x\text{-} y$). The laser pulse dimensionless field $a_0$ ranges from $0.1$ to $0.5$, covering the range of under/overcritical laser pulse powers, and the corresponding power $P/P_{\rm cr,rel}=0.7-4.7$ (peak intensities range from $6\times 10^{16}$ to $4\times 10^{17}$ W/cm$^2$). The uniform plasma locates between $x=5\, \rm \mu m$ and $3$ mm, which is about $2.4$ times the Rayleigh length $L_R =\pi w_0^2/\lambda$. The plasma is comprised of Maxwellian electrons with $T_e=10$ eV and immobile single-charged ions. The electron density is $5\%$ of the critical density $n_{\rm cr} = m_e \omega_0^2/4 \pi e^2$. The simulation box dimension is $200 \lambda \times 100 \lambda$ with the numerical resolution of 16 grid nodes per $\lambda$. Both longitudinal and transverse fastest SMI modes fit in the box. The boundary conditions are outflow for both axes, unless mentioned specifically. The number of particles per cell is 10 per species. We use a moving window simulation setup, so the simulation window starts moving with the primary laser pulse group velocity $c \cdot \sqrt{1 - n_e/n_{\rm cr}}$ right after the laser pulse reaches the $2/3$ of the simulation box. This allows us to simulate up to $10$ ps in order to track the evolution of the laser pulse envelope within the multiple e-folding times of FRS and SMI instabilities.

To find out how the frequency detuning changes the propagation process, we perform a scan on the second laser pulse frequency. The corresponding laser pulse wavelength varies from $1\, \rm \mu m$ to $0.6\, \rm \mu m$. In case of these runs, we separate the total laser power equally into two pulses, one of them with $\lambda =1\, \rm \mu m$, and another one with smaller wavelength. Besides that, we scan on the energy partition between two pulses with $\lambda = 1\, \rm \mu m$ and $0.7\, \rm \mu m$ (frequency detuning $\Delta \omega/\omega_{\rm pe} \approx 1.91$) and three pulses with equally redistributed energy with $\lambda =1\, \rm \mu m$, $0.75\, \rm \mu m$, and $0.6\, \rm \mu m$ (corresponding to $\Delta \omega/\omega_{\rm pe} \approx 1.5$). We verify our results with the 32 nodes per micron grid resolution for a few runs with and without the frequency detuning. Auxiliary 1D and 2D simulations with periodic boundary conditions are also conducted in order to check the importance of the side scattering in the power propagation problem.

We measure the power losses in the process of the laser pulse propagation as the ratio of the electromagnetic energy left in the box to the initial energy of the laser pulse/pulses. Even though some energy of the laser pulse may be converted into other forms of the radiation (i.e. the energy is not actually dissipated, but propagates with group velocity slower/faster than moving window speed), we aim at the propagation of the significant peak power with the least amount of the pulse power lost in order to have an efficient further interaction. 
While this may not be required by all applications, laser ion acceleration above the 100 MeV threshold \cite{Higginson2018} and plasma-based laser amplification \cite{MF2020} will benefit from larger laser pulse powers delivered at the target surface. 

\section{Results}

\subsection{Single laser pulse propagation}

\begin{figure}
    \centering
    \includegraphics[width=\linewidth]{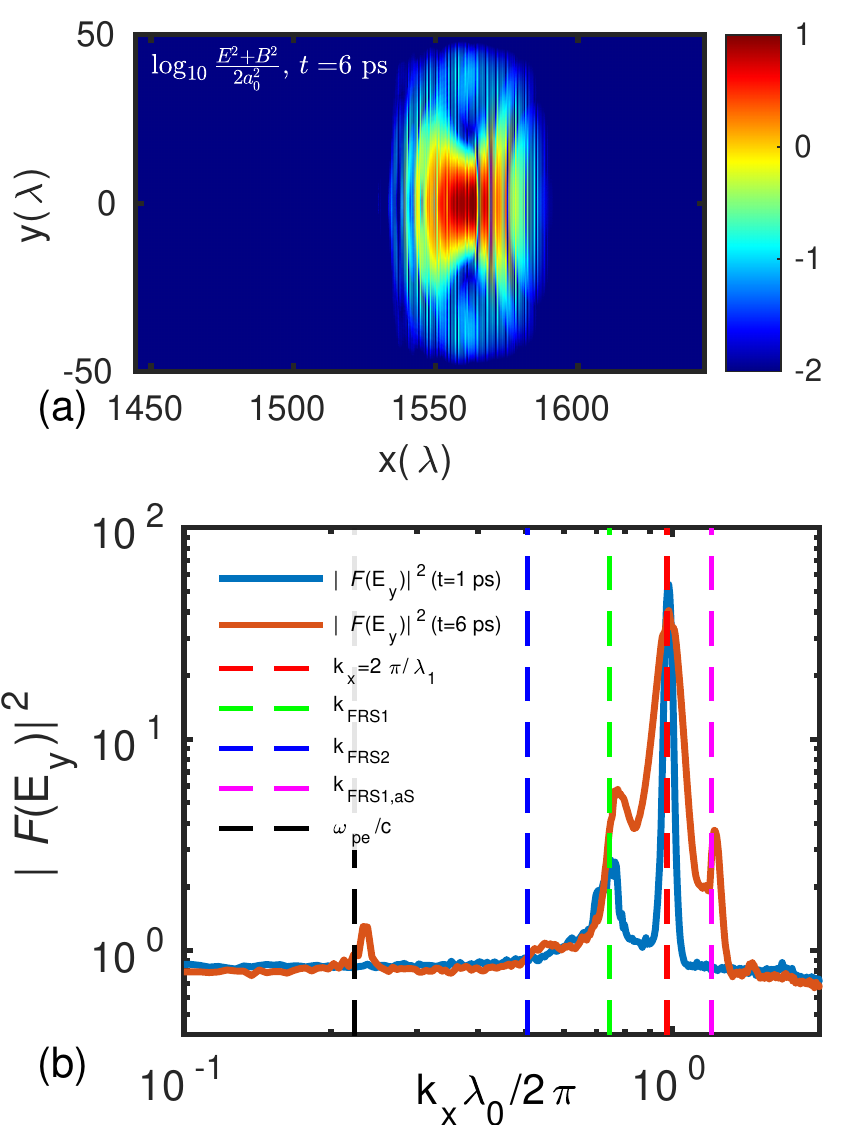}
    \caption{(a) Normalized EM energy distribution and (b) Fourier power spectrum of the electric field $E_y$ for $a_0 = 0.21$ and $n_e/n_{\rm cr}=0.05$ at $t=1$ ps and $6$ ps.}
    \label{fig:sim1}
\end{figure}

First, let us discuss the simulation of the single undercritical laser pulse propagation in the underdense plasma. The considered run has $a_0 \approx 0.21$ (peak intensity $I=6\times 10^{16}$ W/cm$^2$) and $n_e/n_{\rm cr}=0.05$. The corresponding laser power $P/P_{\rm cr,rel}\approx 0.7$ and thus filamentation instability is not developed. 

Figure~ \ref{fig:sim1}(a) shows the electromagnetic energy density distribution of the laser pulse at $t=6$ ps and Fig.~\ref{fig:sim1}(b) shows the transversely averaged longitudinal Fourier power spectrum of the electric field $E_y$ at $t = 1$ ps and $6$ ps. We see both longitudinal and transverse modulations of the pulse envelope in Fig.~\ref{fig:sim1}(a). Both SMI and FRS show their signatures in the Fourier power spectra shown in Fig.~\ref{fig:sim1}(b). Here, we present longitudinal power spectrum of the laser pulse at the initial stage and after $>10$-e-folding growth of FRS instability (blue and brown line, respectively). Vertical lines represent various high-order FRS wavenumbers for the same parameters. The input laser pulse begins with a peak at $k_{\rm med} = \omega_0/c \sqrt{1-{\omega_{\rm pe}^2}/{\omega_0^2}}$. After $\sim 1$ ps, the first-order FRS peak is developed near $k_\mathrm{FRS,1}$. The second-order FRS peak is also shown as a small bump near $k_\mathrm{FRS,2}$. Two other peaks are seen at the plasma wave wavenumber $k_{\rm pe}= \omega_{\rm pe}/c$ and anti-Stokes sideband $k_{\rm FRS,-1} = \sqrt{(\omega_0+\omega_{\rm pe})^2-\omega_{\rm pe}^2}/c$. FRS peaks eventually grow to the order of the primary laser pulse peak height. The spectrum becomes flattened at the later stages, which may be attributed to longitudinal SMI instability~\cite{FilamentationReview}. The resulting power loss after the interaction (at $10$ ps) is $15\%$ and is approximately the same for 1D and 2D simulations with periodic and outflow transverse boundary conditions.

\begin{figure}
    \centering
    \includegraphics[width=\linewidth]{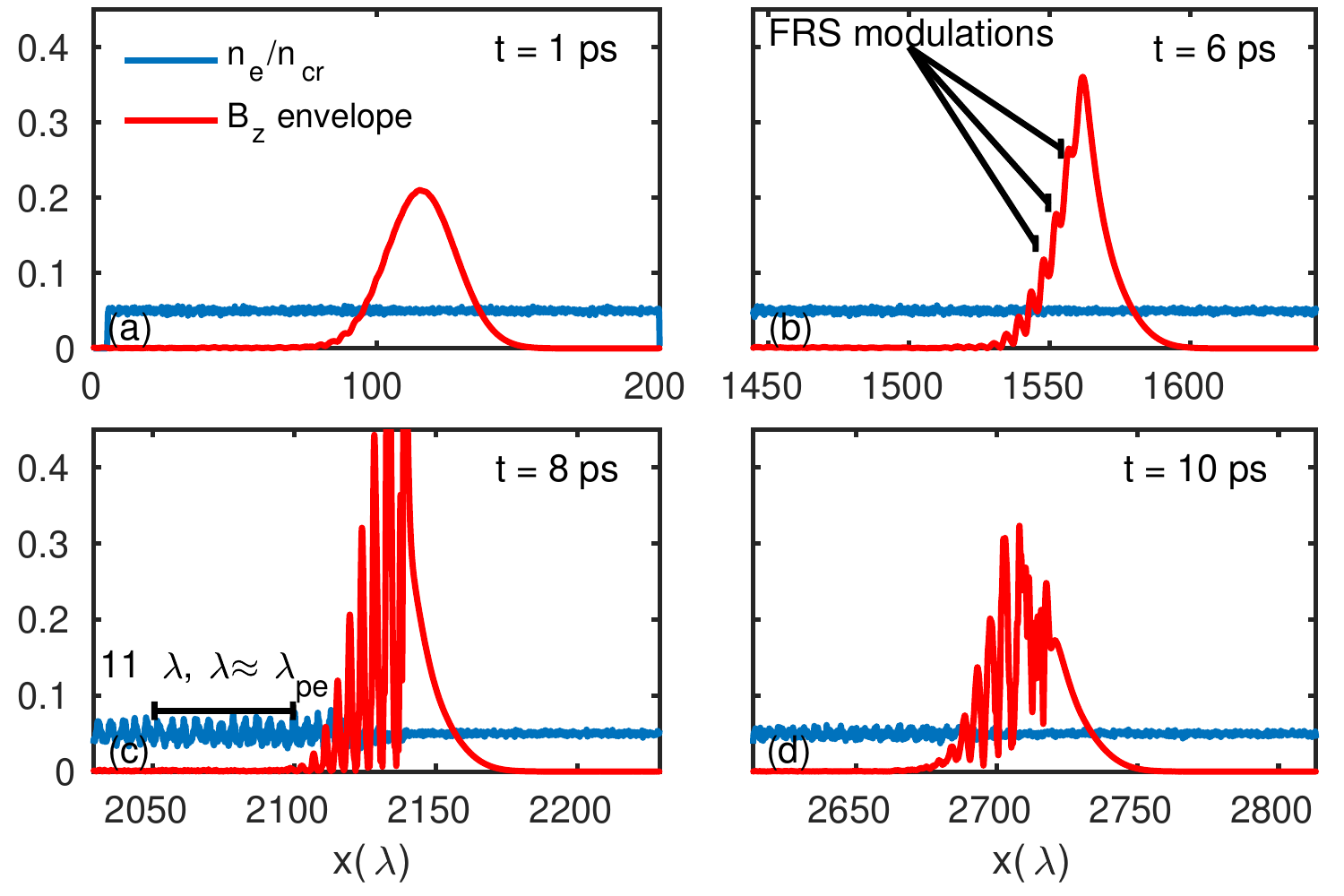}
    \caption{Envelopes of the laser pulse $B_z$ field in units of $m_e \omega_0 c /e$ (red curve) and plasma density (blue curve) with $P/P_{\rm cr,rel}=0.7$ at different times.}\label{fig:enlossmech}
\end{figure}

To analyze the mechanisms of laser power loss, we show four snapshots of the laser pulse envelope and the plasma density, Figure~\ref{fig:enlossmech}. Fig.~\ref{fig:enlossmech}(a) shows the initial laser pulse envelope at $t=1$ ps. At $t=6$ ps, FRS instability begins to modulate the laser pulse tail as shown in Fig.~\ref{fig:enlossmech}(b). The simulation time corresponds to $N_{\rm e, FRS} \approx 15$ e-folding growth. The FRS sideband overlaps with the primary laser pulse, causing a $1.5$ times larger amplitude. At $t=8$ ps, Fig.~\ref{fig:enlossmech}(c) shows strong perturbation of the pulse envelope and periodic structure of the plasma density (blue line). The period of these structures is $\approx 4.35 \lambda$, which is very close to $\lambda_{\rm pe} = c /\omega_{\rm pe} \approx 4.47 \lambda$. Development of the plasma wave peak is also seen in the spectrum shown in Fig.~\ref{fig:sim1}(b). Since the Stokes FRS sidebands have smaller group velocities $v_{\rm gr,FRS1} = c \cdot \sqrt{1- n_e/n_{\rm cr}\cdot \omega_0^2/\omega_{\rm FRS}^2} \approx 0.83 c$ than the primary laser pulse $v_{\rm gr,1} = c \cdot \sqrt{1-n_e/n_{\rm cr}} \approx 0.97 c$, they, together with the plasma wave, flow out from the left boundary of the simulation box and their energies are lost. For the simulation box length of $200\, \mu \rm m$, they will be able to leave the moving window in $\approx 4.6$ ps, as can be seen by comparing Fig.~\ref{fig:enlossmech}(b) and (d). A qualitatively similar behavior is observed for overcritical pulses as well, though both FRS and SMI happen much faster and lead to larger losses of power from the simulation box via FRS. It is also worth mentioning that even though we observe an excitation of the plasma waves in the wake of the laser pulse, the power, according to the Manley-Rowe relations~\cite{DZF008}, is no more than $\omega_{\rm pe}/\omega_{\rm FRS} \approx 30 \%$ in comparison to FRS photons.

\subsection{Frequency detuning of the laser pulse into equal energy pulses}

Next, we show how two copropagating laser pulses with a frequency detuning can suppress the power loss. We separate the total power into two laser pulses with a frequency detuning $\Delta \omega/\omega_{\rm pe}$, following \cite{Kalmykov2006,Kalmykov2008,Kalmykov2009}. In these works, it was shown that the frequency detuning $\Delta \omega/\omega_{\rm pe} > 1$ helps to manipulate the pulse focusing effect, allowing for the more steady process of the laser pulse propagation by avoiding catastrophic self-focusing. While our figure of merit and laser-plasma parameters are different from ones in \cite{Kalmykov2006,Kalmykov2008,Kalmykov2009}, it turned out that the same method can also suppress the FRS growth rates and corresponding power losses.

\begin{figure}
    \centering
    \includegraphics[width=\linewidth]{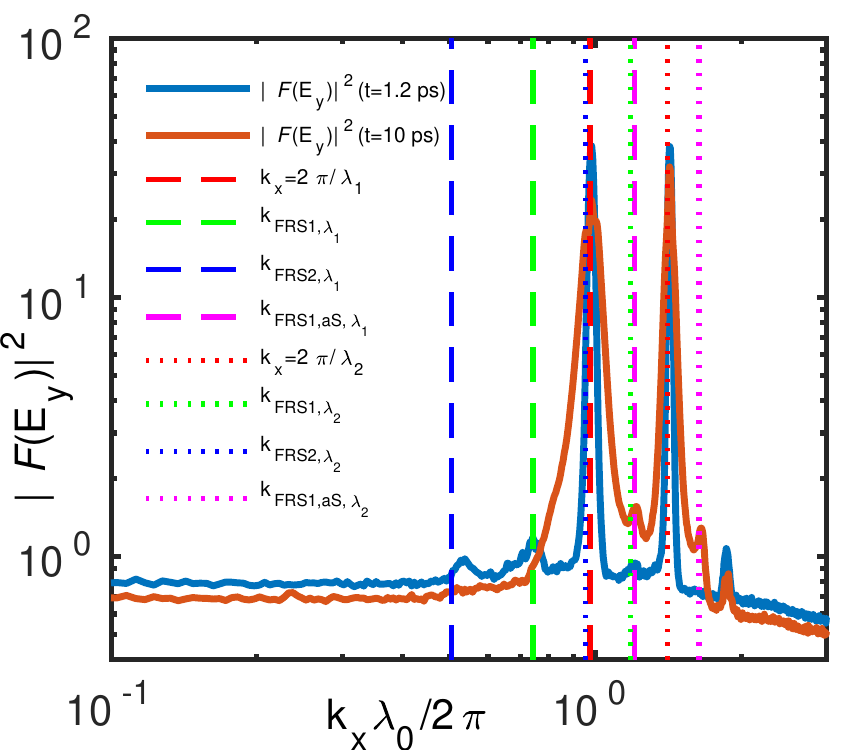}
    \caption{Spectra of the two-color laser pulse before and after propagating through underdense plasma of length $L_x$ with $n_e/n_{\rm cr}=0.05$ for $t=10$ ps. The detuning is $\Delta \omega/\omega_{\rm pe}\approx 1.91$.}
    \label{fig:twocolor}
\end{figure}

Figure \ref{fig:twocolor} shows the evolution of the power spectrum for two-color laser pulse system with the total power $P/P_{\rm cr,rel} \approx 0.7$ and the detuning $\Delta \omega/\omega_{\rm pe}\approx 1.91$. In this case, the percentage of propagated energy is almost 100\%. We see that there is only a slight broadening of spectrum peaks corresponding to $\lambda_1=1\, \mu \rm m$ and $\lambda_2 = 0.7\, \mu \rm m$. FRS peaks do not develop significantly. Comparing with the spectrum of a single laser pulse with the same $P/P_{\rm cr,rel}$ and $n_e/n_{\rm cr}$ in Fig.~\ref{fig:sim1}(b), the FRS peaks developed in the two-color scheme case are at least $5$ times lower.

\begin{figure*}
    \includegraphics[width=\linewidth]{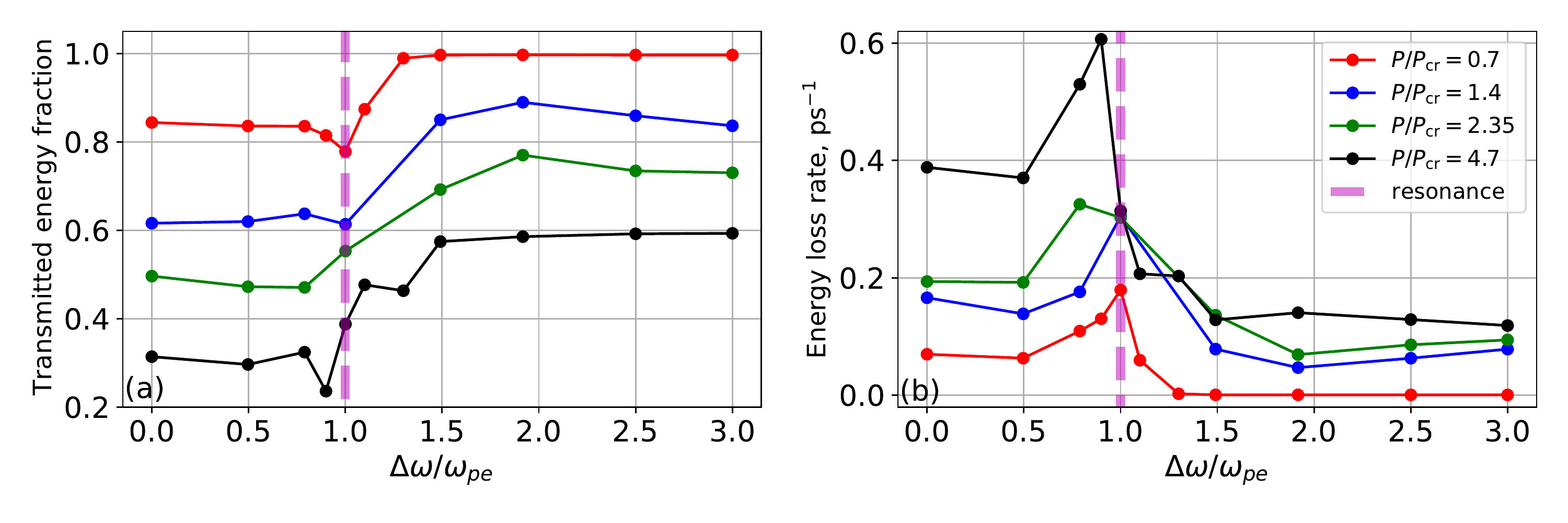}
    \caption{(a) transmitted energy vs frequency detuning $\Delta \omega/\omega_{\rm pe}$; (b) peak energy loss rate vs $\Delta \omega/\omega_{\rm pe}$ for equal energy laser pulses with $n_e/n_{\rm cr}=0.05$. For $\Delta \omega/\omega_{\rm pe} > 1$ the power transmission is clearly more efficient, as the power loss rate is significantly suppressed.}
    \label{fig:domegascan}
\end{figure*}

Finally, we show in Fig.~\ref{fig:domegascan} the fraction of transmitted energy and energy loss rate for different frequency detuning. The curves represent different total laser pulse power $P/P_{\rm cr,rel}\approx 0.7,1.4,2.35,4.7$. The frequency detuning spans from $\Delta \omega/\omega_{\rm pe}=0$ to $3$. The plasma density is $n_e/n_{\rm cr}= 0.05$. The left panel shows the fraction of the injected energy propagated through the plasma slab of length $L_x$; the right panel shows the energy loss rate from the moving window. It clearly demonstrates the function of frequency detuning. For small detunings $\Delta \omega/\omega_{\rm pe} < 1$, there is no significant suppression of power loss. For $|\Delta \omega/\omega_{\rm pe} -1| \ll 1 $, FRS is resonantly driven to worsen the power propagation efficiency in comparison to a single pulse case. However, $\Delta \omega/\omega_{\rm pe}>1$ shows up to $50 \%$ efficiency increase in transmitted energy fraction.

\subsection{Frequency detuning of the laser pulse into unequal energy pulses}
The possible explanation may be formulated in terms of the FRS growth rates - recalling that $\gamma_{\rm FRS} \propto \sqrt{I} \lambda^{3/2}$, we may argue that propagating power with shorter wavelength is more efficient since it allows to suppress FRS development. Even further suppression of the FRS may be achieved by separating the total energy (= total intensity) into pulses with different wavelengths. Auxiliary 1D simulations and 2D simulations with periodic transverse boundary conditions suggested that the side scattering, if present, is not a major contributor to the total power loss.

\begin{figure}
    \includegraphics[width=\linewidth]{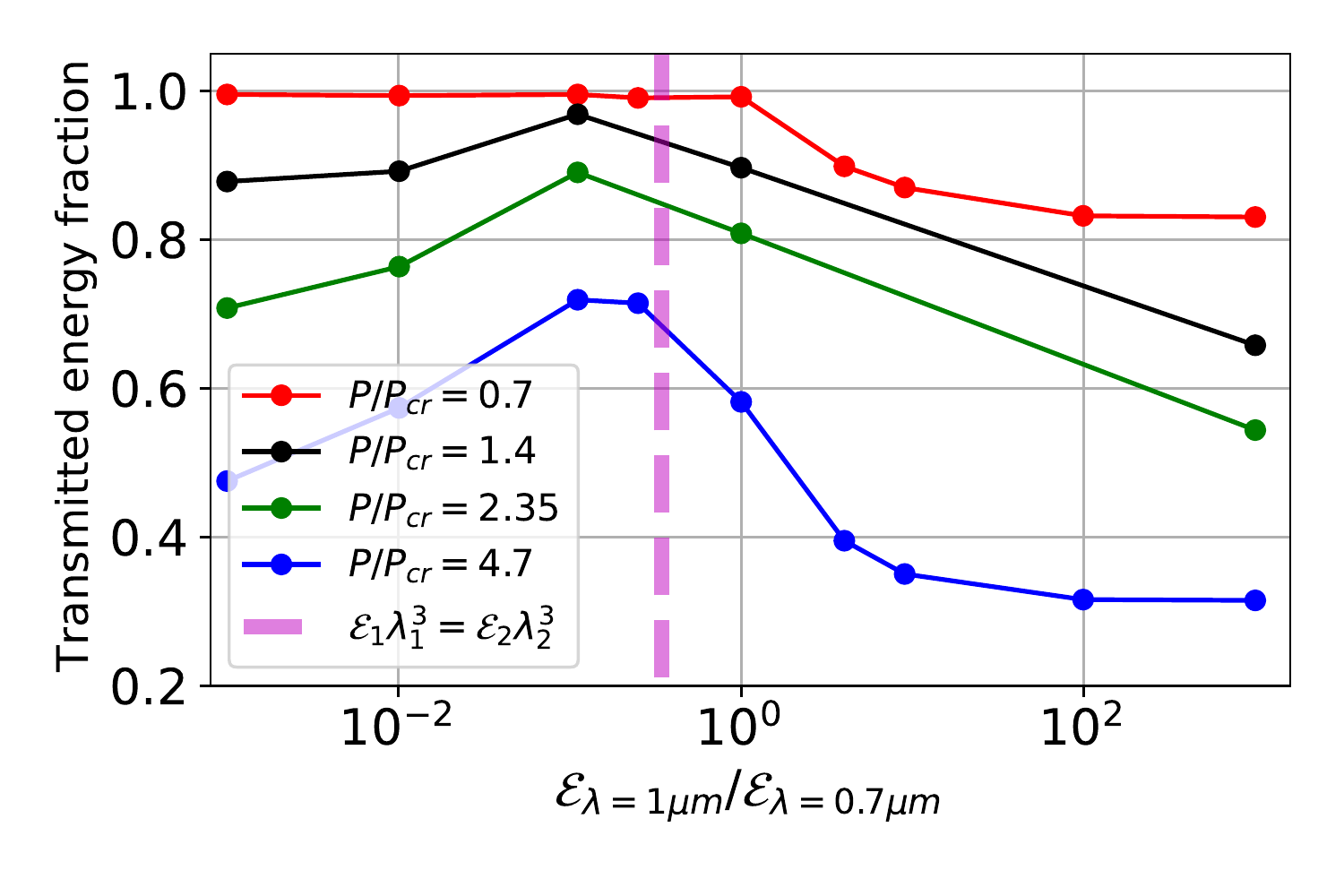}
    \caption{Transmitted energy vs energy partition ratio for simulations with $\Delta \omega/\omega_{\rm pe} \approx 1.91$. The dashed magenta line demonstrates a theoretically predicted optimum.}
    \label{fig:domegascan4}
\end{figure}

We perform a scan on energy ratio between two pulses, ranging from 0.01 to 100, including simulations with the whole energy in one of the wavelengths. Figure~\ref{fig:domegascan4} summarizes the results of the scan. The curves show an optimal energy distribution ratio at which the power losses are minimized. Nontrivially, the optimal energy distribution ratio is not $1$, but near $\sim 0.2$. 

\begin{figure}
    \centering
    \includegraphics[width = \linewidth]{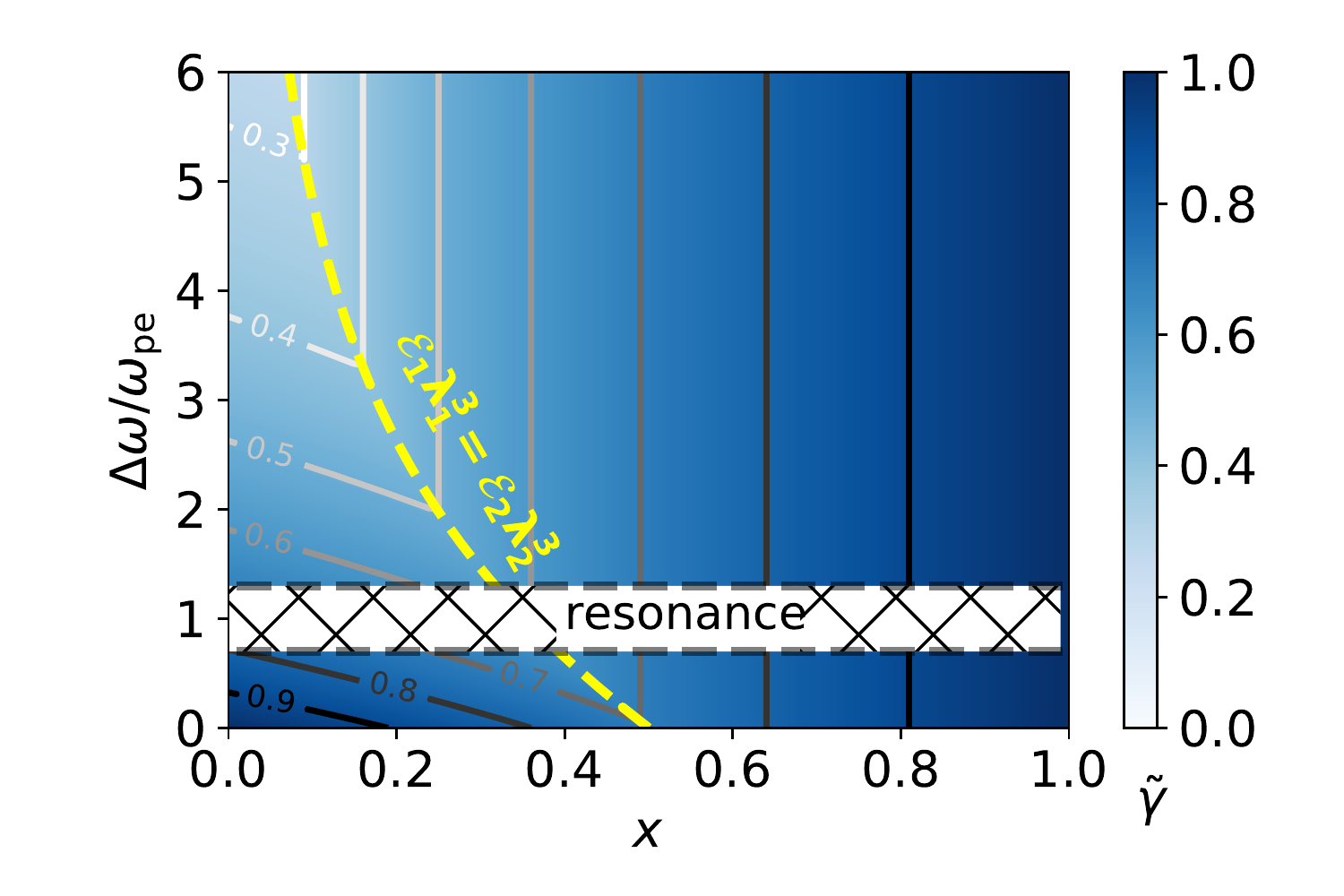}
    \caption{Normalized two-color laser FRS growth rate, $\Tilde{\gamma}$,  as the function of energy partition ratio $x$ and frequency detuning $\Delta \omega/\omega_{\rm pe}$. The hatched region corresponds to the FRS resonance $\Delta \omega/\omega_{\rm pe }\approx 1$, and hence should be avoided. The yellow dashed line represents an optimal regime for the FRS suppression.}
    \label{fig:sup}
\end{figure}

Since the dominating power loss factor is FRS, the nontrivial optimization of the energy distribution ratio for the maximum power transmission can be explained by analyzing the FRS growth rates. Assuming the two frequency components grow independently, the observed two-color laser FRS growth rate is the maximum growth rate of FRS for each wavelength,
\begin{equation}
    \gamma_{\rm FRS}^{2c} = {\rm max} \left(\gamma_{\rm FRS}^{\lambda_1},\gamma_{\rm FRS}^{\lambda_2}\right).
    \label{FRS2crate}
\end{equation}
We normalize it to the single-color laser FRS growth rate, $\gamma_{\rm FRS}^{1c}$, and define
\begin{align}
    \Tilde{\gamma} &\equiv {\rm max} \left(\frac{\gamma_{\rm FRS}^{\lambda_1}}{\gamma_{\rm FRS}^{0}},\frac{\gamma_{\rm FRS}^{\lambda_2}}{\gamma_{\rm FRS}^{0}}\right) \nonumber \\
    &=  {\rm max} \left(x^{1/2}, (1-x)^{1/2}\frac{\lambda_2^{3/2}}{\lambda_1^{3/2}}\right),
    \label{FRS2crate2}
\end{align}
where $x=\mathcal{E}_1/\mathcal{E}_{tot}$ is the ratio of the energy of the first pulse to the total energy of the laser system. Figure~\ref{fig:sup} shows the dependence of the optimal two-color laser FRS growth rate $\Tilde{\gamma}$ on $x$ and frequency detuning $\Delta \omega/\omega_{\rm pe}$. It is seen that the optimal energy distribution starts from $0.5$ at negligible detuning $\Delta\omega$ and decreases at larger $\Delta\omega$. 
Note that the analysis assumes independent FRS for the two frequency components. This assumption is violated at $\Delta\omega=\omega_{\rm pe}$ when the two components resonantly excite FRS. The minimum of $\Tilde{\gamma}$ can be found by equalizing both terms in brackets in Eqn.~(\ref{FRS2crate2}). It will lead to the optimization condition 
\begin{equation}
    \mathcal{E}_1 \lambda_1^3 = \mathcal{E}_2 \lambda_2^3.
    \label{opt}
\end{equation}
The yellow line shows an optimal direction in ($x,\Delta \omega/\omega_{\rm pe}$) space. We plot the dashed magenta line in Fig.~\ref{fig:domegascan4} to show the agreement with the numerical results.

In the similar fashion, it is possible to use more than two pulses for a more efficient power transfer. We performed a few preliminary runs with three pulses separated between each other by at least $\Delta \omega/\omega_{\rm pe} \approx 1.5$. We saw that using three pulses further increases the power transmission efficiency, in agreement with $\Tilde{\gamma}$ scaling of $1/\sqrt{N_{\rm p}}$, where $N_{\rm p}$ is the number of pulses. The energy partition can be optimized among multiple pulses for the best efficiency. Generalization of the approach described here for $N_{\rm p} \gg 1$ and ultimately getting into the incoherent pulse regime is of interest as well~\cite{EM2017}, but will be addressed in the separate work.

\section{Summary and Discussion}

In summary, we discussed how to optimally transmit laser power using a multi-color system through the cold uniform underdense plasma slab by avoiding FRS and filamentation instabilities. 
Using 2D PIC simulations, we identified the primary role of FRS in power losses. 
We further demonstrated how frequency detuning suppresses the FRS power losses. 
The filamentation instability acts only as a secondary factor by modulating the pulse envelope, which leads to modifications in local FRS growth rates. 
We showed that the frequency detuning for the efficient power transmission should be $\Delta \omega/\omega_{pe}> 1$, and the resonant regime with $\Delta \omega \approx \omega_{pe}$ should be avoided. 
By considering the unequal energy partition between detuned pulses and varying the total number of pulses, we found that the frequency detuning has an optimal energy ratio between two pulses, thus verifying the importance of the pulse interplay for the efficient power transmission. 
Using more than two pulses further improves the transmission efficiency.
The extension to multiple pulses suggests the use of an entirely incoherent laser pulse to even further increase the power transmission efficiency~\cite{EM2017}.

Let us compare our results with the results from Kalmykov et al.~\cite{Kalmykov2006,Kalmykov2008,Kalmykov2009}, where the frequency detuning approach was also used, but aimed at suppression of the filamentation instability. Both our and their studies conclude that FRS and longitudinal SMI play a dominant role at the later stages of pulse evolution after the pulse traveled for an extended distance $X>L_R$. It was claimed in~\cite{Kalmykov2008} that the electromagnetic cascading (i.e. the development of higher order FRS modes) cannot be neglected starting from $X\approx 3L_R/8$ for $\Delta \omega/ \omega_{\rm pe} \approx 2$ and equal energy partition case \cite{Kalmykov2006}. The corresponding Stokes FRS modes are indeed seen in our simulations at $t\approx 1$ ps (see Fig 4). The laser energy depletion reported in \cite{Kalmykov2009} is in reasonable agreement with our simulations with $P/P_{\rm cr}=0.7, \, \Delta \omega / \omega_{\rm pe} \approx 1.3$ and $P/P_{\rm cr}=1.4, \, \Delta \omega / \omega_{\rm pe} \approx 1.5$. The overall higher laser energy depletion in our simulations apparently is due to higher $\omega_{\rm pe}/\omega_0$ value than in \cite{Kalmykov2009}. The discrepancy in the most dominant instability at the initial stage of the laser-plasma interaction can be explained by invoking the criteria for the dominance of FRS and SMI instabilities [Eqn. (6)]: using parameters of simulations in \cite{Kalmykov2006,Kalmykov2008,Kalmykov2009} ($P=55 \, \rm TW$, $\tau_L=1.3$ ps, $n_e=5.65\cdot 10^{17}$ cm$^{-3}$, $\lambda=0.8\, \mu \rm m$), we get that in their case, SMI is dominant over FRS ($\Gamma \approx 0.022$), and for parameters of our simulations it is the opposite ($\Gamma \geq 1.8$).

The two-color laser scheme was applied in laser wakefield acceleration \cite{Li2019b} and is proposed to be used for electron-positron plasma generation \cite{Chen2019}, among other applications (see refs in \cite{Li2019b}). The results obtained in our work broaden the applicability of this scheme, highlighting the benefits of multi-color laser systems in terms of an efficient laser power transmission through medium-to-low underdense plasma. 

\section*{Acknowledgements} 
This work was supported by NNSA DENA0003871 and AFOSR FA9550-15-1-0391. 
The EPOCH code was developed as part of the UK EPSRC funded projects EP$/$G054940$/$1. 
The simulations were performed using computational resources at the TIGRESS high-performance computer center at Princeton University.

\section*{Data availability}
The data that support the findings of this study are available from the corresponding author upon reasonable request.

\end{document}